\documentclass{llncs}
\usepackage[T1]{fontenc}
\usepackage{epsfig}
\usepackage{graphicx}
\usepackage{amsmath}
\usepackage{multicol}
\usepackage{xy}
\usepackage[utf8]{inputenc}
\usepackage{algorithm}
\usepackage{algorithmicx}
\usepackage[noend]{algpseudocode}

\newcommand{\intc}{\ensuremath{\mathit{int}}}
\newcommand{\extc}{\ensuremath{\mathit{ext}}}

\algdef{SE}{Block}{EndBlock}%
   [1]{\textproc{#1}}%
   {}
\algtext*{EndBlock}%

\author{Jakub Łącki \and Piotr Sankowski}
\institute{Institute of Informatics\\
 University of Warsaw\\
Warsaw, Poland
}

\title{Min-cuts and Shortest Cycles in Planar Graphs in $O(n \log \log n)$ Time}

\begin{document}

\maketitle

\begin{abstract}
We present a deterministic $O(n \log \log n)$ time algorithm for finding shortest cycles and minimum cuts in planar graphs.
The algorithm improves the previously known fastest algorithm by Italiano {\it et al.} in STOC'11 by a factor of $\log n$.
This speedup is obtained through the use of dense distance graphs combined with a divide-and-conquer approach.
\end{abstract}

\section{Introduction}
In this paper we study the minimum cut and shortest cycle problems in planar graphs.
The {\it minimum cut problem} is to find the cut with minimum capacity, whereas the {\it shortest
cycle problem} is to find the cycle with minimum total length. These two problems are actually equivalent, since a shortest cycle corresponds to a minimum cut in the dual graph. Moreover, the size of the minimum cut is equal to the weighted edge-connectivity of the graph.
In this paper when presenting the algorithms we only talk about the problem of finding shortest cycle keeping in mind that the min-cut problem can be solved using this reduction.

In general graphs the minimum cut can be found in $O(m \log^3 n)$ randomized time as
shown by Karger~\cite{karger-00}, or in $O(mn+n^2 \log n)$ deterministic time as given by Nagamochi and Ibraki~\cite{nagamochi-ibraki-90}. On the
other hand, the shortest odd cycle can be found in $O(nm)$ time~\cite{itai-rodeh-78}, whereas the shortest even length cycle
can be found in $O(n^2)$ time~\cite{yuster-zwick-97}.

In the case of planar graphs these two problems have attracted considerable attention in recent years. Even in the case of the unweighted graphs
these problems are interesting. However, one needs to keep in mind that the duals of unweighted graphs are not longer unweighted.
Eppstein~\cite{eppstein-99} was the first one to show how to find cycles of constant weight in $O(n)$ time. This result was later on improved
by Alon, Yuster and Zwick~\cite{alon-yuster-zwick-97}, who have shown an $O(n)$ time algorithm for finding shortest cycles of length $\le 5$. This actually implies a linear time algorithm for minimum cuts in planar unweighted graphs, as the sizes of the minimum cuts are at most $5$. On the other hand, the fastest algorithm for finding the shortest cycle in unweighted graph (also called the {\it girth} of the graph), was given by Weimann and Yuster~\cite{weimann-yuster-09}. This very recent algorithm works in $O(n\log n)$ time.

In the case of weighted planar graphs the fastest algorithm was proposed in this year and works in $O(n \log n\log \log n)$ time~\cite{italiano-11}. This algorithm is obtained by simply joining the $O(n \log^2 n)$ time divide-and-conquer approach of Chalermsook {\it et al.}~\cite{chalermsook-04} with a faster $O(n \log \log n)$ time max-flow algorithm given in~\cite{italiano-11}. Here we show an even further improvement by showing an $O(n\log \log n)$ time algorithm for both minimum cut and shortest cycle problem in weighted planar graphs. This not only improves over the result in~\cite{italiano-11} but also over the result of Weimann an Yuster~\cite{weimann-yuster-09} for unweighted graphs. The minimum cut problem is related to {\it minimum $st$-cut problem}, where we need to find minimum cut that separates $s$ from $t$. Up until the paper of Italiano {\it et al.}~\cite{italiano-11} the fastest known algorithm worked in $O(n \log n)$ time~\cite{frederickson-87}. You may notice that the approach of Chalermsook {\it et al.}~\cite{chalermsook-04} results in a $\log n$ complexity gap between min-cut and min $st$-cut problems. In this paper we actually show how to close this gap, and we believe that the techniques introduced here will be useful as well when faster min $st$-cut algorithm are developed in the future. 

In addition to the static results, we give a dynamic algorithm for computing the minimum cut size and the shortest cycle length in planar graphs. The processes updates and answers queries in $O(n^{5/6} \log^{5/3}n)$ time and is the first known dynamic result that is able to handle weighted edges and arbitrary edge connectivity. The only previously known exact dynamic algorithm with update time sublinear in $n$ was able to maintain the information about polylogarithmic edge connectivity in $O(\sqrt{n})$ time per update~\cite{thorup-01}. For the history of this problem we refer you to the description in~\cite{thorup-01}. 

This paper is organized as follows. In next section we give a summary of the techniques developed in the previous papers that we use. In Section~\ref{section-chalermsook-algorithm} we recall the Chalermsook {\it et al.}~\cite{chalermsook-04} algorithm. Our first algorithm that works in $O(n \log n)$ time is given in Section~\ref{n-log-n-algorithm}. This algorithm actually builds a part of the main result of this paper that is given in Section~\ref{section-computing-the-girth}. The dynamic algorithm is given in the final section of this paper. 

\section{Preliminaries}
\label{section-preliminaries}
For a graph $G = (V,E)$, we denote the set of its vertices and edges by $V(G)$ and $E(G)$ respectively.
Additionally, if $G$ is planar, $F(G)$ denotes the set of its faces.
If $C$ is a cycle in a planar graph, we define its interior and exterior (denoted by $int(C)$ and $ext(C)$) to be the subgraphs embedded inside and outside the cycle, both containing the cycle itself.

\paragraph{Simplifying Assumptions}
We will assume that the graph we work on is both triangulated and has a constant degree. This can be easily achieved $O(n)$ time by
first triangulating the dual graph with zero weight edges and then triangulating the primal graph with infinite length edges using
zigzag triangulations~\cite{BorradaileSW10}.

\paragraph{Graph $r$-division}
Define a \emph{piece} $P = (V_P,E_P)$ of $G$ to be the subgraph of $G$ defined by a subset $E_P$ of $E$.
In $G$, the vertices of $V_P$ incident to vertices in $V\setminus V_P$ are the \emph{boundary vertices} of $P$.
We will denote the set of boundary vertices of piece $P$ by $\partial P$.
Vertices of $V_P$ that are not boundary vertices of $P$ are \emph{interior vertices} of $P$.

We define an \emph{$r$-division} of $G$, to be a division of (the edges of) $G$ into $O(n/r)$ pieces each containing $O(r)$ vertices and $O(\sqrt r)$ boundary vertices. A \emph{hole} is a finite face of $P$ which is not a face of $G$. The following theorem was shown by Italiano \emph{et al.}~\cite{italiano-11}.

\begin{theorem}\label{theorem-r-division}
For a plane $n$-vertex graph, an $r$-division in which each piece has $O(1)$ holes can be found in $O(n\log r + (n/\sqrt r)\log n)$
time.
\end{theorem}
In this paper, when we talk about an $r$-division, we shall assume that it has the form as in Theorem~\ref{theorem-r-division}.

For an $r$-division $\mathcal{P}$, define a {\it skeleton graph} $G_{\mathcal{P}}=(\partial \mathcal{P}, E_{\mathcal{P}})$ to be a graph over the set of border vertices in $\mathcal{P}$. The edge set $E_{\mathcal{P}}$ is composed of infinite length edges connecting consecutive border vertices of each hole.

\paragraph{Dense Distance Graphs}
We use the $r$-division to define a representation for shortest paths in a graph that has similar number of edges, but fewer vertices.
In order to achieve this, we use the notion of dense distance graphs.
If $G$ is edge-weighted, we define the \emph{dense distance graph} of a piece $P$ to be the complete graph on the set of boundary vertices of $P$ where each edge $(u,v)$ has weight equal to the shortest path distance (w.r.t.\ the edge weights) in $P$ between $u$ and $v$.
A dense distance graph for an $r$-division is a set of dense distance graphs of all its pieces.
Observer that it contains $O(\frac{n}{\sqrt{r}})$ nodes and $O(n)$ edges.

Italiano \emph{et al.}~\cite{italiano-11} have used an algorithm by Klein~\cite{klein-05} to compute a dense distance graph for any division.

\begin{lemma}
\label{lemma-dense-graphs}
Given an $r$-division $\mathcal{P}$, its dense distance graph can be computed in $O(n \log r)$ time.
\end{lemma}

\paragraph{Fast Dijkstra}
The dense distance graphs can be used to speed up shortest path computations using Dijkstra's algorithm.
It was shown by Fakcharoenphol and Rao (\cite{fakcharoenphol-06}, Section 3.2.2) that a Dijkstra-like algorithm can be executed on a dense distance graph for a piece $P$ in $O(|\partial P| \log^2 |P|)$ time. Having constructed the dense distance graphs, we can run Dijkstra in time almost proportional to the number of vertices (rather than to the number of edges, as in standard Dijkstra). We use this algorithm in graphs composed of dense distance graphs and a subset $E'$ of edges of the original graph $G=(V,E)$:

\begin{corollary}\label{corollary-fast-dijkstra}
  Dijkstra can be implemented in $O(|E'| \log |V|  +  \sum_{i}|\partial G_i|\log^2 |\partial G_i|)$ time on a graph composed of a set of dense distance graphs $G_i$ and a set of edges $E'$ over the vertex set $V$.
\end{corollary}
\begin{proof}
In order to achieve this running time we use Fakcharoenphol and Rao~\cite{fakcharoenphol-06} data structure for each $G_i$. Moreover, minimum distance vertices from each $G_i$ and all endpoints of edges in $E'$ are kept in a global heap.
\end{proof}

\paragraph{Max Flow Queries}
\label{section-max-flow-queries}
Italiano \emph{et al.}~\cite{italiano-11} have shown an $O(n^{2/3}\log^{8/3}n)$ time dynamic algorithm for computing max-flow values in planar graphs. More generally speaking, they have presented an algorithm that allows the following tradeoffs between preprocessing, update and query times.

\begin{theorem}
\label{theorem-dynamic-shortest-paths}
There exists a data structure that after  $O(n \log r + \frac{n}{\sqrt{r}}\log n)$ preprocessing time, supports:
edge insertions and edge deletions in $O((r+\frac{n}{\sqrt{r}})\log^2 n)$ time;
$s$ to $t$ distance queries in $O((r+\frac{n}{\sqrt{r}})\log^2 n)$ time;
max $st$-flow queries in $O((r+\frac{n}{\sqrt{r}})\log^3 n)$ time,
where $r\in [1,\ldots, n]$.
\end{theorem}

The only information maintained by the algorithm is an $r$-division together with dense distance graphs for all pieces.
If we set $r = \log^8 n$, then the initialization takes $O(n \log \log n)$ time and the query time becomes $O((\log^8 n + \frac{n}{\log^4 n})\log^3 n) = O(\frac{n}{\log n})$.
Hence, we obtain the following static algorithm that will be very useful for us:

\begin{corollary}
\label{corollary-max-flow-queries}
There exists a data structure that after $O(n \log \log n)$ preprocessing time, can compute a max $st$-flow value in $O(\frac{n}{\log n})$ time.
\end{corollary}

\section{Chalermsook {\it et al.} Algorithm}
\label{section-chalermsook-algorithm}
Chalermsook, Fakcharoenphol and Nanongkai~\cite{chalermsook-04} have shown an $O(n \log^2 n)$ algorithm (we call it CFN from now on) for finding minimum cuts in undirected, weighted planar graphs.

The algorithm of Chalermsook~\emph{et al.} uses a divide and conquer approach (given as Algorithm~\ref{alg:CFN}).

\begin{algorithm}[t]
\begin{algorithmic}[1]
\Procedure{CFN Algorithm}{planar graph $G$}
\Block{Reduce:}
\State{replace each degree $2$ vertex $v$ with an edge, whose weight is equal to the sum of weights of edges incident to $v$}
\EndBlock
\Block{Divide:}
\State{find a shortest paths tree $T$}
\State{find an edge $bc$ in $G-T$ such that the cycle $C$ in $T+bc$ contains a constant fraction of faces of $G$}
\State{recurse on $G\cap \intc(C)$ to find the shortest cycle $C_i$ inside $C$}
\State{recurse on $G\cap \extc(C)$ to find the shortest cycle $C_x$ outside $C$}
\EndBlock
\Block{Conquer:}
\State{find shortest cycle $C_c$ that crosses $C$}
\State{return the shortest cycle from $C_e$, $C_x$ and $C_c$}
\EndBlock
\EndProcedure
\end{algorithmic}
\caption{CFN algorithm for finding the shortest cycle in a planar graph $G$}
\label{alg:CFN}
\end{algorithm}

\paragraph{Dividing Step}
The algorithm computes a shortest paths tree using a linear time algorithm by Henzinger~\emph{et al.}~\cite{henzinger-97}. Then it finds a cycle $C$ that divides the graph into two parts, both containing a constant fraction of all faces. The cycle $C$ consists of two shortest subpaths $Q_{ab}$ and $Q_{ac}$, that belong to the tree, and a path $bc$, which goes along a boundary of a face. After that, the algorithm recursively computes the length of the shortest cycles inside and outside $C$. In addition, a shortest cycle that crosses $C$ is computed.

\paragraph{Conquering Step}
Let $Q = q_1q_2\ldots q_k$ be some shortest path in $G$, and let $F_i$ be any face incident to $q_i$, for $i=1$ and $i=k$. The following is
a well know fact (see e.g.,~\cite{reif-81}).
\begin{lemma}
\label{lemma-crossing-shortest-path}
The length of the shortest cycle which crosses $Q$ an odd number of times is equal to the length of the shortest cycle which separates $F_1$ from $F_k$.
For every cycle $C$ that crosses $Q$ an even number of times, there exists a cycle which does not cross $Q$ and is not longer than $C$.
\end{lemma}

Let $F_e$ be a face adjacent to $bc$ that lies inside $C$ and let $F_a$ be a face from outside of $C$ that is adjacent to the first edge of $Q_{ab}$.
The above lemma implies directly the following result due to Chalermsook \emph{et al.}~\cite{chalermsook-04}.
\begin{lemma}
\label{lemma-chalermsook}
The length of the shortest cycle which crosses $C_e$ an odd number of times is equal to the length of the shortest cycle which separates $F_a$ from $F_e$.
For every cycle $C$ that crosses $C_e$ an even number of times, there exists a cycle which is fully contained either in $\intc(C_e)$ or $\extc(C_e)$ and is not longer than $C$.
\end{lemma}
By duality of shortest cycles and minimum cuts, we can find such a shortest cycle using a single maximum flow computation. This can be done in $O(n \log \log n)$ time using a recent algorithm by Italiano \emph{et al.}~\cite{italiano-11}.
In the following, whenever we talk about the shortest cycle crossing some path or cycle, we actually mean the shortest cycle that crosses it an odd number of times.

\paragraph{Reducing Step}
In the reducing step, we remove vertices of degree $2$ by merging their incident edges.
As a result, all vertices have degree at least $3$ and, by Euler's formula, the number of vertices is at most twice the number of faces.
Moreover, each dividing step adds at most one new face, so the total number of faces in every recursion level is bounded by $O(n)$.
The same bound holds for the number of vertices.
There are $O(\log n)$ recursion levels and each requires $O(n \log \log n)$ time. Hence, the overall running time is $O(n \log n \log \log n)$.

\section{An $O(n \log n)$ Time Algorithm}
\label{n-log-n-algorithm}
In this section we show how to obtain a faster algorithm by a simple modification of the CFN algorithm. We present an improved version of the CFN algorithm that still has $O(\log n)$ levels of recursion, but each of them will require $O(n)$ amortized time.

We now run the recursion as follows.
Every $\log \log n$ levels of the recursion, in every branch of the recursion tree we reinitialize the maximum flow algorithm from Corollary~\ref{corollary-max-flow-queries}. Over all levels of the recursion this takes $O(\frac{\log n}{\log \log n} n \log \log n) = O(n \log n)$ time. Within $\log \log n$ levels following the initialization, we issue at most $2^{\log\log n} = \log n$ maximum flow queries to each maximum flow structure. As we have observed, this requires only linear time. Hence, all maximum flow computations require $O(n \log n)$ time.

The data structure is not recomputed in every step, but only from time to time. Hence, a query for a shortest cycle separating two faces is answered using the structure for larger part of the graph. However, this does not affect the final result. When the graph has some additional vertices and edges, the length of the shortest cycle can only decrease. Moreover, every cycle we find is a valid cycle in the original graph, so the length of the shortest cycle in the whole graph is computed correctly.

The CFN algorithm runs in $O(n \log n)$ time when we exclude time needed for maximum flow computations. Here, we have shown how to perform all maximum flow computations in $O(n \log n)$ time, thus reducing the running time of the whole algorithm from $O(n \log n \log \log n)$ to $O(n \log n)$.

\section{An $O(n \log \log n)$ Time Algorithm}
\label{section-computing-the-girth}
We show that the algorithm of Chalermsook \emph{et al.}~\cite{chalermsook-04} can be implemented on \emph{dense distance graphs}.
Instead of recursing on the subgraphs, we use the skeleton graph.
The dense distance graphs are kept in a global memory and referred when needed.
We follow the structure of Section~\ref{section-chalermsook-algorithm} and describe how to implement all three steps of the algorithm.
However, here we stop the recursion when the subgraph for recursion contains less than $r$ nodes.
Hence, we require a terminal step that handles such small subgraphs at the end of the algorithm.

The first step in the CFN algorithm is building a shortest paths tree.
We also start by computing a shortest paths tree $T$ in a dense distance graph of an $r$-division.
However, we require $T$ to be \emph{noncrossing}, which means that its every edge can be mapped to an underlying shortest path inside one piece in such a way, that the paths do not cross \footnote{Note that if a piece has holes, this can happen even if edges of $T$ can be embedded in the plane without crossing.}.
We use the linear time algorithm by Henzinger \emph{et al.}~\cite{henzinger-97} for finding shortest paths in a planar graph with nonnegative edge weights.
Then for each piece of the decomposition we map the shortest subpaths connecting border vertices to their corresponding edges in the dense distance graph.
The resulting tree is noncrossing.

The main part of the algorithm is based on a divide and conquer technique.
We start by building an $r$-division and a skeleton graph for the given graph.
We define a \emph{recursion graph}.
This graph will be used to represent the parts of the entire $r$-division that are considered in recursive calls.
Initially this is a skeleton graph.
In every step the graph is divided by intersecting it with an interior and exterior of some cycle.
To represent this process, we insert \emph{division edges} to the recursion graph.
Those edges connect border vertices belonging to one piece.
We never add vertices to the recursion graph, so it has $O(\frac{n}{\sqrt{r}})$ vertices all the time.

A \emph{region} is a subgraph of the recursion graph bounded with division edges.
Every region represents a part of the graph that is processed in one recursive call (see Fig.~\ref{fig:recursion-graph}).

Regions contain faces of two kinds.
Some faces contain parts of the graph that are represented by this region (e.g. the light gray faces of region $A$ in Fig.~\ref{fig:recursion-graph}).
We call those faces \emph{internal faces}.
The rest of faces contains parts of the graph, that are to be processed in other branches of the recursion (e.g. the face of region $A$ which contains region $B$ or the outer face of $A$).
Every face of the graph belongs to exactly one region, but edges and vertices can be in multiple regions.

Note that we can insert a division edge connecting some pair of border vertices  multiple times.
This means that for some region $R$ only the shortest path connecting those two vertices belongs to $R$.
Since we want the recursion graph to be simple, if there are multiple division edges connecting a pair of border vertices within the same piece, we merge them into a single edge.
Such edges can belong to many regions.

Whenever we need to extract the distance between two border vertices from one piece, we check whether they belong to the same internal face of the region or if they are connected with a division edge (or a super edge, which is defined later).
If this is the case, we return length of the appropriate edge from the dense distance graph.
Otherwise, the distance is infinite.

\subsection{Dividing Step}
\label{subsection-dividing-step}

The dividing step of the CFN algorithm finds a cycle that splits the graph into two subgraphs, both containing at most $\frac{2}{3}$ of all faces.
It is a fundamental cycle determined by one edge in a shortest paths tree.
This is what we do as well, but we use the noncrossing shortest paths tree to do it more efficiently.

We find a cycle that is composed of two subpaths $Q_{ab'}$ and $Q_{ac'}$ of the shortest paths tree $T$, two shortest paths $Q'_{b'b}$ and $Q'_{c'c}$ (in the original graph) fully contained within one piece of the decomposition and an edge $bc$.
Those paths form a cycle which cuts the original graph into two pieces, each containing a constant fraction of faces.

We build a planar graph $F$ by taking the union of edges of the skeleton graph $G_{\mathcal{P}}$ and $T$.
$F$ can contain multiple edges between some pairs of vertices.
A spanning tree $S_F$ of its dual can be constructed by taking edges corresponding to edges from the skeleton graph (see Fig.~\ref{fig:graph-f}).


We assign weights to vertices of $S_F$.
A vertex $v$ of $S_F$ corresponds to a face $f$ from $F$.
We define its weight as the number of faces of $G$ which are inside $f$ in their common embedding.
Those values can be computed right after building the shortest paths tree in the beginning.

Denote by $W$ the total weight of all vertices.
The goal is to find a vertex $w$ in $S_F$ such that the total weight of its subtree is at least $\frac{W}{2}$ and the weight of its every child subtree is smaller than $\frac{W}{2}$.
This can be achieved in linear time by walking down the tree (starting from the root) and always choosing the heaviest subtree.

Let us denote by $P$ the piece containing the face corresponding to $w$.
We build a graph $F_P$ in the following way.
First, we take the tree $T$ and, by running Dijkstra's algorithm in piece $P$, we extend $T$ with paths to all internal vertices of $P$, thus obtaining a tree $T_P$.
This requires $O(r \log r)$ time.
The graph $F_P$ consists of the skeleton graph $G_{\mathcal{P}}$, the tree $T_P$ and all edges of $P$, which do not belong to $T_P$.

Again, we build a spanning tree $S_{F_P}$ of the dual graph, by taking edges that do not belong to $T_P$.
The weight of a vertex $v \in V(S_{F_P})$ is the number of faces of $G$ embedded inside $v$.
Thus, total weight of $S_{F_P}$ is equal to the total weight of $S_F$.
In fact, $S_{F_P}$ can be obtained from $S_F$ by splitting some vertices.


Then we find a vertex $w$ in the tree $S_{F_P}$, such that the total weight of its subtree is at least $\frac{W}{2}$ and the weight of subtrees rooted in all its children is smaller than $\frac{W}{2}$.
If we apply the routine describe earlier and break ties the same way as previously, we find a vertex $w$, which corresponds to a face belonging to piece $P$.
This means that the weight of $w$ is equal to $1$ and its degree is $3$.
It is easy to observe that by removing an appropriate edge $e'$ incident to $w$, we can split the tree into two parts which contain a constant fraction of the total weight.

The edge $e$ which is a primal edge corresponding to $e'$ determines a fundamental cycle $C_e$ in $T_P$, which splits the graph into two pieces, both containing a constant fraction of all faces of the original graph.
We now want to cut the graph with this cycle and recurse into two smaller subgraphs.

It remains to show how to carry out the cutting.
We describe how to do it for pieces that contain no holes.
The general case is handled in the appendix.

The cycle $C_e$ consists of shortest subpaths between border vertices (except for the piece $P$).
Hence, for any piece $P'$ other than $P$ shortest paths both in $P' \cap int(C_e)$ and $P' \cap ext(C_e)$ do not cross $C_e$.
This implies that the distances in both parts are preserved, so it suffices to insert one division edge that connects two respective border vertices to the recursion graph.

To cut the piece $P$, we have to use a different approach.
We use Klein's algorithm~\cite{klein-05} to rebuild the dense distance graphs of $P \cap int(C_e)$ and $P \cap ext(C_e)$ in $O(r \log r)$ time, as given by Corollary~\ref{lemma-dense-graphs}.
A division edge is inserted between the border vertices of $P$ connected by $C_e$.

The last step is to cut the noncrossing shortest paths tree $T$.
Since the cut $C_e$ does not cross any edge of $T$, suffices to divide it into $int(C_e) \cap T$ and $ext(C_e) \cap T$.
Note that the shortest path tree $ext(C_e) \cap T$ is rooted in the vertex $a$, which belongs to $C_e$.

\subsection{Conquering Step}
\label{subsection-conquering-step}

In the previous section we have shown how to find a cycle $C_e$, which divides the graph into two subgraphs of roughly the same size.
The cycle $C_e$ can be mapped to a fundamental cycle determined by a single edge $e$ in a shortest paths tree of the original graph.
This means that it consists of two shortest subpaths $Q_{ab}$ and $Q_{ac}$ and the edge $e = bc$. By Lemma~\ref{lemma-chalermsook}, to find the shortest cycle crossing $C_e$ we need a single maximum-flow computation.
Note that the cycle $C_e$ might not be simple, i.e. $Q_{ab}$ and $Q_{ac}$ can share some edges in the beginning, but the lemma still holds.
As discussed in the previous section, we have the $r$-division and dense distance graphs ready in each recursive call.
This allows us to use the algorithm from Theorem~\ref{theorem-dynamic-shortest-paths} to answer max st-flow queries in $O((r + \frac{n}{\sqrt{r}}) \log^3 n)$ time.

\subsection{Reducing Step}
\label{subsection-reducing-step}
In the CFN algorithm in the reduction step we have removed degree $2$ vertices.
This was done in order to bound the total number of vertices in all branches on each level of the recursion.
Here, we would like to do the same.

The total number of vertices in one level of the recursion is the total number of vertices in all regions.
Some vertices can be counted many times, if they belong to multiple regions.
Each face, on the other hand, is in exactly one region.

Consider a region $R$.
Some of its vertices are adjacent to the faces of the recursion graph, which are inside $R$.
It follows that the number of such vertices is equal to the sum of sizes of all faces from the interior of $R$, so there are $O(\frac{n}{\sqrt{r}})$ such vertices in total.

$R$ can also contain some vertices that are not adjacent to any face from its interior.
From among those, we find vertices incident to exactly two division edges.
If we take the part of the original graph corresponding to $R$, those vertices also have degree two.
Therefore, we remove each such vertex and replace its incident edges $e_1$ and $e_2$ with a \emph{super edge}, whose length is the sum of lengths of $e_1$ and $e_2$ (in Fig.~\ref{fig:recursion-graph}, the path from region $A$ going along the boundary of region $B$ is replaced with a super edge).
Note that there can be no vertices incident to $0$ or $1$ division edge in $R$.
For the analysis, let us also get rid of all other vertices of degree $2$ with a a similar procedure.
We obtain a planar graph $R'$ with vertices of degree at least $3$.

The degree bound implies that $|V(R')| = O(|F(R')|)$.
Removing vertices of degree two does not increase the number of faces, so $|V(R')| = O(|F(R)|)$ \footnote{Here $F(R)$ denotes the set of all faces of $R$, not only those from its interior.}.
In every recursion step, we divide one region into two regions with a single cycle.
This increases the total number of faces in all regions by a constant.
Since there are $O(\frac{n}{\sqrt{r}})$ recursion steps, we conclude that $|V(R')| = O(\frac{n}{\sqrt{r}})$, so there are $O(\frac{n}{\sqrt{r}})$ vertices not adjacent to an interior face in any region.

\begin{corollary}
\label{vertices-in-regions}
The total number of vertices in all regions in every recursion level is $O(\frac{n}{\sqrt{r}})$.
\end{corollary}

Observe that the same argument allows us to bound the number of super edges with $O(\frac{n}{\sqrt{r}})$.

\subsection{Terminal Step}
Each recursive call for region with $k$ vertices consists of the following steps:
\begin{enumerate}
\item Find a cycle $C_e$ that, when mapped to the original graph, divides it into two parts containing constant fraction of faces.
This requires $O(k + r \log r)$ time.
\item Compute the length of the shortest cycle separating two given faces.
This step runs in $O((r+k)\log^3 k)$.
\item Insert a division edges corresponding to $C_e$ to the recursion graph and divide the shortest path tree.
\end{enumerate}

Hence, the total running time of each call is $O((r+k)\log^3 k)$.
We run the recursion as long as $r \leq k$.
This implies that the total cost of each step would be dominated by the summand depending on $k$.
Since we start with a graph with $O(\frac{n}{\sqrt{r}})$ vertices, the recursion takes $O(\frac{n}{\sqrt{r}} \log^4 n)$ total time.

If in any recursive call $r > k$, we abandon the recursion and use a different approach for finding the shortest cycle within the current region $R$.
We will refer to such recursive calls as \emph{terminal} recursive calls.

For each region $R$ in from a terminal recursive call we need to compute the part of the original graph, which corresponds to it.
We find a graph $G_R$, which represents the part corresponding to $R$, in a compressed way.
Namely, some paths composed of vertices of degree $2$ are replaced with one edge.
The process of finding graphs $G_R$ is described in the appendix.

\begin{lemma}
\label{terminal-graphs}
All graphs $G_R$ can be computed in $O(n \log r)$ time.
The sum of $|V(G_R)|$ over all regions $R$ is $O(n)$.
\end{lemma}

It remains to find the length of the shortest cycle within each $G_R$.
In order to do that, we use the $O(n \log n)$ algorithm for computing the girth from section \ref{n-log-n-algorithm}.
The following lemma is necessary to bound the running time of the terminal calls.

\begin{lemma}
For each $R$, $|V(G_R)| = O(r^2)$.
\end{lemma}

\begin{proof}
By the definition of the terminal recursive call, $|V(R)| < r$, so $G_R$ can contain at most $r$ vertices that are border vertices of some pieces.
All other vertices lie inside the interior faces of $R$.
Each such face is fully contained within one piece of the $r$-division, so it contains at most $O(r)$ vertices.
Since there are at most $O(r)$ faces in $R$, we conclude that there are $O(r^2)$ vertices in $G_R$ in total.\qed
\end{proof}

From the above lemmas, it follows that running the running time of the $O(n \log n)$ algorithm on all $G_R$ is
\begin{align*}
\sum_R O(|V(G_R)| \log |V(G_R)|) & \leq \sum_R O(|V(G_R)| \log r^2)\\
       & = O(\log r) \sum_R |V(G_R)| = O(n \log r).
\end{align*}

Thus, the whole algorithm requires $O(\frac{n}{\sqrt{r}} \log^4 n) + O(n \log r)$ time.
Setting $r = \log^8n$ yields an $O(n \log \log n)$ algorithm.

\section{Dynamic Shortest Cycle}
\label{section:dynamic-shortest-cycle}

In this section we show how to use the ideas introduced in the previous section to construct a dynamic algorithm for finding minimum cuts in planar graphs. We show how to maintain a planar graph $G$ with positive edge weights under an intermixed sequence of the following operations:
{\em insert$(x,y,c)$} add to $G$ an edge of weight $c$ between vertex $x$ and vertex $y$, provided that the embedding of $G$ does not change;
{\em delete$(x,y)$}  delete from $G$ the edge between vertex $x$ and vertex $y$;
{\em shortest-cycle} return the length of the shortest cycle in $G$.

\subsection{Shortest Cycles through Given Edge}
In our algorithm we use a dynamic data structure that supports the following operations:
{\em insert$(x,y,c)$} add to $G$ an edge of weight $c$ between vertex $x$ and vertex $y$, provided that the embedding of $G$ does not change;
{\em delete$(x,y)$}  delete from $G$ the edge between vertex $x$ and vertex $y$;
{\em shortest-cycle$(x,y)$} return the length of the shortest cycle that includes edge $(x,y)$.

The existence of such dynamic algorithm is implied by Theorem~\ref{theorem-dynamic-shortest-paths}.
\begin{lemma}
\label{lemma:dynamic}
Given a planar graph $G$ with positive edge weights, we can insert edges, delete edges and report the length of a shortest cycle that includes given edge in $O(n^{2/3}\log^{5/3} n)$ worst-case time per operation.
\end{lemma}
\begin{proof}
For supporting updates we simply use Theorem~\ref{theorem-dynamic-shortest-paths} for $r=n^{2/3}\log^{2/3} n$. When we need to answer a query for an edge $(x,y)$ we:
delete edge $(x,y)$;
ask for the shortest distance from $x$ to $y$ -- this plus the length of $(x,y)$ is the shortest cycle length;
reinsert edge $(x,y)$.
Hence, the answer to a {\em shortest-cycle$(x,y)$} query can be computed using three operations of the original data structure.\qed
\end{proof}

\subsection{Data structures and Updates}
In our dynamic algorithm we maintain two data structures from Theorem~\ref{theorem-dynamic-shortest-paths}:
{\em structure $A$} for $r=n^{1/3}$; {\em structure $B$} for $r=n^{2/3}\log^{2/3} n$.

Initialization of both structures requires $O(n \log n)$ time.
Additionally, in the beginning we compute the length of the shortest cycle fully contained within each piece used by structure $A$.
This also runs in $O(n \log n)$ time.

The edge updates are handled using Theorem~\ref{theorem-dynamic-shortest-paths}, i.e., for each piece in the decomposition we maintain the dense distance graph. Additionally, in structure $A$ we find a shortest cycle contained fully inside the piece, which the inserted or deleted edge belongs to. This does not increase running time of the algorithm even when we use the $O(n \log n)$ time algorithm from Section~\ref{n-log-n-algorithm}. The update time is then $O(n^{5/6}\log^2 n)$.

Our algorithm for answering queries follows the lines of the $O(n \log \log n)$ algorithm. However, instead of using the $r$-division for $r=\log^8n$ we use the $r$-division given by structure $A$ for $r=n^{1/3}$. In previous section we have essentially shown that for polylogarithmic $r$ we obtained polylogarithmic speed up for the CFN algorithm. By taking polynomial $r$ we are able to obtain a polynomial speed up of CFN. For $r=n^{1/3}$ the speed up will be by a factor of $n^{1/6}$ and the running time of our algorithm will be $O(n^{5/6}\log^3 n)$. Nevertheless there are some technical differences between the static and dynamic algorithm which are included in Appendix~\ref{appendix-answering-dynamic-queries}. The main difference is that we might need to divide the graph using cycles defined by several non-tree edges. In order to find shortest cycles crossing many non-tree edges we use Lemma~\ref{lemma:dynamic} applied to structure $B$.  

\bibliographystyle{plain}
 \bibliography{girth-planar-near-linear}

\appendix
\begin{figure}
\centerline{\psfig{file=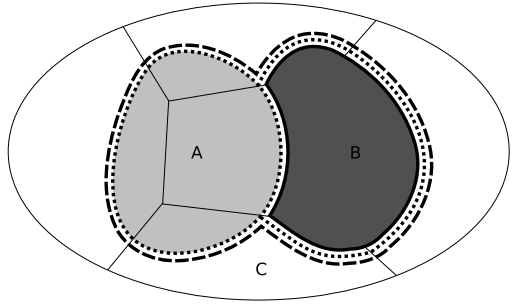, scale=0.8}}
\caption{The recursion graph with three regions $A$, $B$ and $C$.
The interiors of regions are colored light gray, gray and white, respectively.
Each edge is marked with two or three strokes to represent boundaries of regions going through this edge.
In particular, the path along the boundary of region $B$ belongs to all three regions.}
\label{fig:recursion-graph}
\end{figure}

\begin{figure}
\centerline{\psfig{file=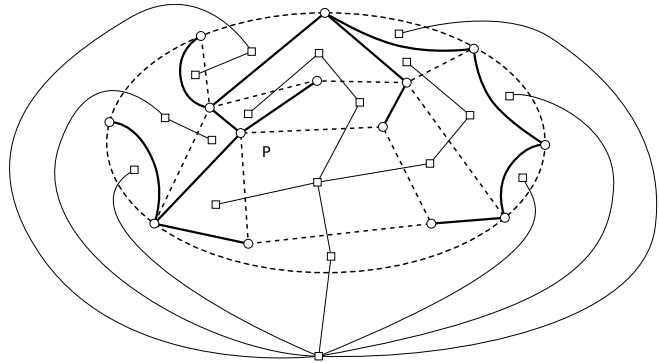, scale=0.7}}
\caption{Graph $F$ and the spanning tree $S_F$.
Vertices of the skeleton graph are marked with circles and its edges with dashed segments.
Bold edges are the edges from the shortest paths tree $T$.
The spanning tree of the dual graph ($S_F$) consists of the square vertices connected with thin solid arcs.}
\label{fig:graph-f}
\end{figure}

\begin{figure}
\centerline{\psfig{file=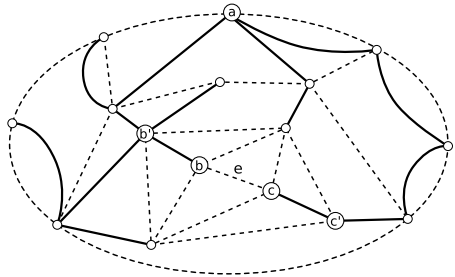, scale=0.7}}
\caption{Graph $F_P$.
Vertices are marked with circles and its edges with dashed segments.
Bold edges are the edges from the shortest paths tree $T_P$.
The spanning tree of the dual has been omitted for readability.}
\label{fig:graph-f-p}
\end{figure}

\section{Dealing with holes}

In this section we show how to deal with some issues that might arise when the $r$-division contains holes.
By Theorem \ref{theorem-r-division}, we can assume that there are is at most a constant number of holes in every piece.

\subsection{Intersecting pieces with a cycle}
\begin{figure}
\centerline{\psfig{file=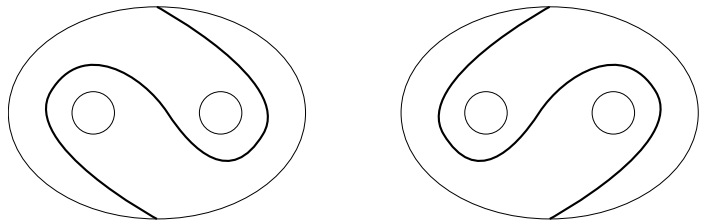}}
\caption{Two ways of embedding a division edge (marked in bold), which correspond to the same division of the set of holes.}
    \label{not-homotopic}
\end{figure}

In the dividing step of the algorithm we intersect some pieces with the interior and exterior of a cycle $C_e$.
This is done implicitly, by inserting division edges, but if a piece has holes, each edge can be embedded in various ways.
For our algorithm it suffices if the edge is inserted in such a way, that it partitions the set of holes into two parts in the same way as the corresponding shortest path.
This does not imply that the edge is homotopic to the shortest path (see Fig.~\ref{not-homotopic}).

This cycle $C_e$ consists of edges from dense distance graphs and a path going within one piece.
We cut this piece by tracking the cycle and rebuilding dense distance graphs on its both sides in $O(r \log r)$ time.
It remains to describe how to intersect a piece $P$ with a cycle consisting of edges from the dense distance graph.

We consider the subgraph of the skeleton graph induced by border vertices of $P$.
It has one connected component that is composed of border vertices that do not belong to any hole.
Let us call this component a \emph{frame} of $P$.

We take a subpath of $C_e$ that connects two vertices of frame of $P$ (call this subpath $C_P$).
$C_e$ can go through $P$ multiple times, but we consider all such subpaths one at a time.

First, we cut all holes recursively, so we can replace every part of $C_P$, which is contained within one hole, with a single edge.
$C_P$ cuts this graph into two parts.
For each hole we need to compute which part it belongs to.
This is easy if a hole is cut into two pieces with $C_P$, but a bit more complex in other cases.

We extend the initial computation of the dense distance graph.
Assume that there are $l$ holes $H_1, H_2, \cdots, H_l$.
For each hole $H_i$ we fix a path in the dual graph from the outer face to the hole.
This path corresponds to a curve $K_{H_i}$ that connects an edge from the frame of $P$ with an edge connecting two border vertices of $H_i$.
Such a curve determines a set of edges $E_{H_i}$ which are crossed by it.

After that, for each edge of the dense distance graph we want to compute whether its underlying shortest path contains an even or odd number of edges from $E_{H_i}$.
In order to do that, when computing the dense distance graphs, we modify edge lengths.
For an edge $e$ we replace its length $d$ with a tuple $(d, h_1, h_2, \cdots, h_l)$, where $h_i \in \{0, 1\}$ is equal to $1$ iff $e$ belongs to $E_{H_i}$.
The tuples are compared in lexicographic order, which assures that the shortest paths are not affected with this operation.
Nor is the running time, as there is a constant number of holes.
As a result, we know how many edges from each $E_{H_i}$ are crossed by every edge of the dense distance graph.

\begin{figure}
\centerline{\psfig{file=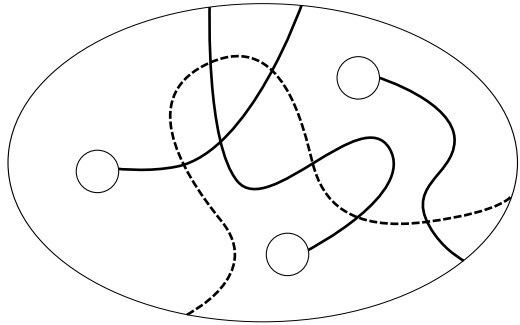}}
\caption{Determining, which side of $C_P$ does every hole belong to. $C_P$ is marked with a dashed line, curves $K_{H_i}$ are solid.}
    \label{embedding-division-edges}
\end{figure}

Once we know that, when we cut the piece $P$ with $C_P$, for every hole $H$ we know the parity of the number of crosses of $C_P$ and $K_H$.
We also know the circular ordering of the beginning of $K_H$ and the first and the last vertex of $C_P$.
This allows us to determine which side of the cut does every hole belong to.
Assume that $H_i$ is not crossed by $C_P$.
Then, it is on the same side of $C_P$ as the beginning of $K_{H_i}$ if and only if the number of times $C_P$ crosses $K_{H_i}$ is even.
This is because every time $K_{H_i}$ crosses $C_P$, it switches to a different side of $C_P$ (see Fig.~\ref{embedding-division-edges}).
Hence, we can embed the division edge in any way, which is coherent with the computed partition of holes.

Lastly, we have to cut the dense distance graph of $P$ into two parts.
We simply insert one division edge to the recursion graph and assume that the distances inside each part are equal to distances inside $P$.
This is in general incorrect for pieces with holes, as some shortest path might cross $C_P$ many times.
Hence, it might occur that the real distance between two vertices is bigger than we have assumed.
This, however, does not affect our final result, as every distance corresponds to some valid path in the original graph, contained within the same piece.

The lengths of the edges are used only in minimum $st$-separating cycle computation in the conquering step ($s$ and $t$ are faces), carried out with Italiano \emph{et al.} algorithm~\cite{italiano-11}.
If we take the computed cycle and map every edge of the dense distance graph to the shortest path, which determines the length of this edge, we obtain a closed walk.
We have to show that this walk bounds some nonempty set of faces in the graph.
This follows easily from the algorithm, as it works with a graph composed of pieces containing $s$ and $t$ and dense distance graphs of all other pieces.
Hence, the walk has to separate $s$ from $t$, which means that it has nonempty interior.
It can be shorter than the shortest cycle contained within the currently considered part of the graph, but is surely not shorter than the globally shortest cycle.

\subsection{Additional preprocessing from the algorithm by Italiano \emph{et al.}}
In the presence of holes, the dynamic algorithm by Italiano \emph{et al.}~\cite{italiano-11} performs additional preprocessing.
For each piece $P$ and each pair of its holes $h$ and $h'$ (in this section, the outer face of the piece is also considered a hole) we fix a path with some structural properties, that connects these holes.
Then, a we compute the dense distance graph for the piece after making an incision along this path.
This dense distance graph is used when we ask for the shortest cycle that separates faces contained in $h$ and $h'$.

As the divide and conquer algorithm proceeds, we also have to cut those dense distance graphs.
This is, however, done implicitly by updating the recursion graph.
As previously shown in the previous section, the distances after performing the cut in this way might be in general invalid, but every distance corresponds to some valid path that is contained within the same piece and does not cross the incision.
This guarantees that every cycle we find is a valid cycle in the entire graph and is not longer than the shortest cycle in the currently processed part of the graph.

\section{Proof of Lemma~\ref{terminal-graphs}}
\label{appendix-scanning}
In order to reconstruct all $G_R$, we need to map every division edge to the underlying subpath in the original graph. There are two types of division edges.
Edges of the first type correspond to paths in piece $P$, which are computed explicitly in the division step.
We can store all such paths in a global memory and use them for division later on.

The division edges of the second type correspond to shortest paths connecting border vertices of $P$. We will first split each piece in the graph using edges of the second type. Then we will use the first type edges to split the resulting regions into smaller parts.

Note that there are at most $O(\sqrt{r})$ such edges as they do not cross. Due to our assumption that the graph has constant degree, a path of length $l$ from one piece can be reconstructed in $O(l)$ time with Klein's algorithm~\cite{klein-05}. If $a$ is a border vertex and $b$ is an arbitrary vertex, the algorithm can report the first edge of the shortest $b$ to $a$ path in constant time, provided that the graph has constant degree. However, as these paths may share many edges we need to be more careful and assure that we will not scan each edge too many times. In the worst case it could take $O(r^{3/2})$ time. 

We consider division edges of $P$ one by one. Take the division edge $e$ and find the shortest path $\pi$ corresponding to it. We split the piece $P$ into parts on both sides of $\pi$. Both parts include vertices and edges of $\pi$. Moreover, we include into both parts a compact description of $\pi$ given by a balanced binary tree containing edges of $\pi$, and mark the edges on $\pi$ as scanned.

Division edges do not cross so each time we take new edge $e=(a,b)$ it will be contained in one part of $P$. Moreover, the scanned edges lie only on the boundary of this part and as a consequence each edge belongs to one compact representation of some path that goes on the boundary. Observe that a vertex on the boundary may belong to at most two compact representations. If it belongs to two compact representations it is the end vertex in both these path. Hence, each time we detect that a vertex belongs to two compact representations we can join them to form one. This takes $O(\log r)$ time.

To find the shortest path $\pi$ corresponding to a division edge we start a linear walk using Klein's algorithm. When we find an edge that was already scanned we look on the compact representation of the path $\rho$ which the edge belongs to. Using binary search we find an edge $(x,y)$ on $\rho$ that can belong to the shortest path from $a$ to $b$. To check if $(x,y)$ can be on the path from $a$ to $b$, we use Klein's data structure to check if the next edge on the path from $x$ to $b$ is equal to $(x,y)$. Both parts obtained by splitting along $\pi$ will contain $\rho \cap \pi$ on their border. However, because $\rho$ was already on the border in one of these parts, all internally shared vertices of $\rho\cap \pi$ will have degree $2$. Observe that we can represent the subpaths of $\rho \cap \pi$ in this part as a single edge of length equal to the length of $\rho \cap \pi$. Hence, the compact representation of subpaths of $\rho \cap \pi$ will be needed only in the other part. Extracting the compact representation for each shared subpath takes $O(\log r)$ time.

This splitting procedure means that each edge will be contained in at most two compact representations of the path. Hence, the total size of the data structures used is bounded by the size of the piece $P$. To bound the running time of this procedure we need to bound the number of binary searches performed in piece $P$. We join consecutive compact representations that lie on the border so each binary search needs to end with an unscanned edge. Hence, we will not perform more binary searches then there are edges in the piece. Moreover, the number of joins and splits performed on compact representations is bounded by the number of binary searches. The cost of each operation is $O(\log r)$, so the time needed for splitting each piece using the edges of the second type is $O(r \log r)$. This gives $O(n\log r)$ total time over the whole graph. On the other hand, splitting according to the first type edges requires $O(n)$ time in total.

If we excluded the time needed for performing binary searches, the above algorithm would run in $O(n)$.
Consequently, the total number of vertices in all graphs $G_R$ is $O(n)$.

\section{Answering Dynamic Queries}
\label{appendix-answering-dynamic-queries}
In order to answer queries we execute the divide and conquer algorithm both on structure $A$ and $B$. As previously, the description of the algorithm follows the structure of Section~\ref{section-chalermsook-algorithm}.

The main data structure for us is structure $A$ and we use its $r$-division as we did in Section~\ref{section-computing-the-girth}. The regions and division edges are defined with respect to structure $A$ as well. As previously, we start the recursion with one region which is given by the skeleton graph. However, we continue the recursion till we obtain regions containing a single cycle.

\paragraph{Dividing Step}
In each recursive call we first find a shortest path tree $T$ in the dense distance graph for the $r$-division in structure $A$. Such tree can be constructed in $O(n^{5/6}\log^2 n)$ time using Corollary~\ref{corollary-fast-dijkstra}.

We build a planar graph $F$ by taking the union of edges of the skeleton graph $G_{\mathcal{P}}$ and $T$.
Next, we build a spanning tree $S_F$ of its dual by taking edges corresponding to edges from the skeleton graph.
In contrary to Section~\ref{section-computing-the-girth}, we assign to all vertices of $S_F$ weights $1$.
The total weight is $W = |F(S_F)|$.
Now, we find a vertex $w$ in $S_F$ such that the total weight of its subtree is at least $\frac{W}{2}$ and the weight of subtrees rooted in all its children is smaller than $\frac{W}{2}$.
Let $d_0, d_1, d_2, \cdots, d_k$ be the edges incident to $w$, listed in circular ordering, where $d_0$ is the edge connecting $w$ to its parent.
We aim to find a sequence of consecutive children, whose total weight is between $\frac{W}{4}$ and $\frac{3W}{4}$.
Since the weights are smaller than $\frac{W}{2}$, there exists some $l \in \{1, 2, \cdots, k\}$ such that the total weight of subtrees connected to $w$ with $d_1, d_2, \cdots, d_l$ satisfies the required inequalities.

\begin{figure}
\centerline{\psfig{file=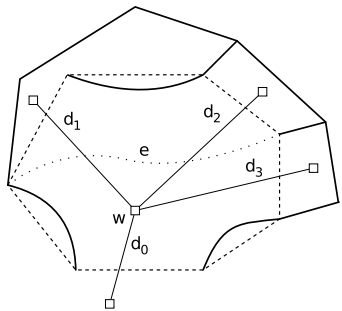, scale=0.8}}
\caption{The face of the skeleton graph, which $w$ belongs to. The boundary of the face is marked with a dashed line. The edges of the shortest paths tree are drawn in bold. A part of the dual tree $S_F$ is marked with thin solid segments. }
    \label{dynamic-dividing}
    \end{figure}

Let $d_0', d_1', \cdots, d_l'$ be primal edges corresponding to $d_0, d_1, \cdots, d_l$.
We select those two endpoints of $d_1$ and $d_l$ which are closer to $d_0$ in the circular ordering around the face $w$ and connect them with an edge $e$ (see Fig.~\ref{dynamic-dividing}).

The edge $e$ determines a fundamental cycle $C_e$ in $T$.
$C_e$ consists of two shortest subpaths $Q_{ab}$ and $Q_{ac}$ and the edge $e = bc$.
Let $C_P$ be the cycle of infinite weight edges that bounds $P$.
Vertices $b$ and $c$ split $C_P$ into two paths $P_1$ and $P_2$.

\begin{figure}
\centerline{\psfig{file=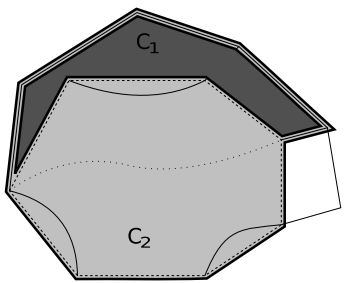, scale=0.8   }}
\caption{Cycles $C_1$ and $C_2$. The interior of $C_1$ is dark gray, whereas the interior of $C_1$ is light gray.}
    \label{dynamic-cycles}
    \end{figure}

The paths $Q_{ab}$ and $Q_{ac}$ when combined with $P_1$ or $P_2$ give two cycles.
Let $C_1$ be the smaller one, and let $C_2$ be the bigger one (see Fig.~\ref{dynamic-cycles}).
These two cycles divide the graph into three regions $\intc(C_1)$, $P = \extc(C_1) \cap \intc(C_2)$ and $\extc(C_2)$. We only need to recurse our procedure onto $\intc(C_1)$ and $\extc(C_2)$. Note that the shortest cycle contained fully inside $P$ is maintained by our dynamic algorithm and we can simply look it up.

It remains to show how to split our structures $A$ and $B$ when recursing.
Consider a piece $P'$ and its dense distance graph $D_{P'}$. In Section~\ref{subsection-dividing-step} we argued that for all pieces but one we do not need to recompute dense distance graphs. However, now we cannot afford to do so at all. Instead we will simply use the dense distance graph for the original non-divided piece, i.e., we will set $D_{P' \cap \intc(C_1)} = D_{P'} \cap \intc(C_1)$ and $D_{P' \cap \extc(C_2)} = D_{P'} \cap \extc(C_2)$. Note that
now distances in, e.g., $D_{P'\cap \intc(C_1)}$ may be different from distances in $P'\cap \intc(C_1)$. However, they can only be shorter in the case when the corresponding shortest path in $P'$ leaves $P'\cap \intc(C_1)$. As we argued in Section~\ref{n-log-n-algorithm}, this will not change the result of the algorithm, i.e., the shortest cycle found only might become shorter, but it exists in the original graph. The distances in the dense distance graphs might not correspond to distances in the pieces, but they do correspond to paths in the original graph.

This implies that we can actually use cycle $C_1$ and $C_2$ to split both structures $A$ and $B$ into two parts.
Our selection of these cycles guarantees that both $\intc(C_1)$ and $\extc(C_2)$ contain a constant fraction of faces of $F$.

\paragraph{Conquering Step}
We are left to show how to find shortest cycles that cross $C_1$ or $C_2$. Note that such cycles need to cross one of $Q_{ab}$, $Q_{ac}$, $P_1$ and $P_2$. In order to find shortest cycles crossing path $Q_{ab}$ and $Q_{ac}$ we find faces incident to their ends and apply Lemma~\ref{lemma-crossing-shortest-path}. On the other hand, a cycle that crosses $P_1$ or $P_2$ has to contain an edge incident to $C_P$.
By the assumption that the graph has constant degree there are at most $O(n^{1/6})$ such edges, because $P$ contains $O(n^{1/6})$ border nodes.
For each such edge $E$ we dispatch a query shortest-cycle$(e)$ to the structure $B$. This takes $O(n^{1/6} n^{2/3}\log^{5/3}) = O(n^{5/6}\log^{5/3}n)$ time in total.

\paragraph{Reducing Step}
In this query answering algorithm we use the same reducing procedure for structures $A$ and $B$ as we used in Section~\ref{subsection-reducing-step}. Hence, Lemma~\ref{vertices-in-regions} implies that the total number of vertices in all regions in every recursion level is $O(n^{5/6})$ for structure $A$ and $O(n^{2/3}\log^{2/3}n)$ for structure $B$.
In every recursive call we divide the graph $F$ into two parts, both containing a constant fraction of faces.
However, the shortest paths tree is recomputed at the beginning of every recursive call.
This does not affect the depth of the recursion, as $F$ contains the same number of faces, regardless of what the shortest paths tree looks like.
This implies that there are $O(\log n)$ levels of the recursion and the total running of the procedure is dominated by structure $A$ and is $O(n^{5/6} \log^3 n)$.

\paragraph{Terminal Step}
We run the recursion until the current region represents a single cycle.
In the terminal step, it remains to compute its length.

\end{document}